\documentclass{article}

\usepackage[main,final]{neurips_2026}
\usepackage{CJKutf8}
\usepackage{aas_macros}

\usepackage[utf8]{inputenc} 
\usepackage[T1]{fontenc}    
\usepackage{hyperref}       
\usepackage{url}            
\usepackage{booktabs}       
\usepackage{amsmath}        
\usepackage{amsfonts}       
\usepackage{amssymb}        
\usepackage{nicefrac}       
\usepackage{microtype}      
\usepackage{xcolor}         
\usepackage{graphicx}       
\usepackage{xspace}

\newcommand{\pz}{photo-$z$\xspace}
\newcommand{\pzs}{photo-$z$'s\xspace}

\title{Tabular Foundation Models Dramatically Outperform Bespoke Photometric Redshift Estimators}

\author{%
   Biprateep Dey$^{1,2}$\thanks{contact: \texttt{biprateep@pitt.edu}} \ \&  \ Joshua S.~Speagle (\begin{CJK*}{UTF8}{gbsn}沈佳士\end{CJK*})$^{1}$ \\ \\
$^{1}$University of Toronto \quad $^{2}$Vector Institute\\
}

\begin{document}

\maketitle

\begin{abstract}
  Estimating galaxy redshifts (a proxy for distance) from imaging data alone, also known as photometric redshifts (\pzs), is essential for studies of extragalactic astrophysics and cosmology. However, data-driven estimators rely on spectroscopic training datasets, which are scarce and systematically biased toward bright and easily observed galaxies.
  In this work, we show that tabular foundation models (TFMs), whose synthetic pretraining amounts to a learned prior over regression problems, can outperform bespoke estimators using only in-context learning. Evaluating on a widely used benchmark dataset, we find that TFMs predicting full probability distributions in a single forward pass systematically outperform bespoke estimators across both representative and biased training data regimes, matching or exceeding baselines trained on over $40{,}000$ galaxies with just $1{,}000$ training context examples.
  We also show that the TFM advantage is largest when little to no calibration data are available, and that it persists even when the bespoke baselines are supplemented with up to $10{,}000$ representative calibration spectra. This shows that TFMs offer a sample-efficient framework for upcoming imaging-only astronomical surveys, where observational selection bias can be mitigated simply by including a small representative calibration sample in context{\renewcommand{\thefootnote}{\fnsymbol{footnote}}\setcounter{footnote}{1}%
\footnote{Code available as a Python package: \url{https://github.com/biprateep/lazy-tfm}}}.
\end{abstract}

\section{Introduction}

Determining galaxy distances is fundamental to extragalactic astrophysics and cosmology. Because spectroscopy is observationally expensive, multi-band imaging (photometry) vastly outpaces spectroscopic data volume, requiring astronomers to infer photometric redshifts (\pzs) from imaging alone. Upcoming surveys like the Vera C.~Rubin Observatory's Legacy Survey of Space and Time \citep[LSST;][]{Ivezic2019} will observe billions of galaxies and rely critically on \pzs. Existing estimators broadly divide into physical template-fitting and data-driven machine learning models trained on spectroscopic sets \citep[see, e.g.,][for reviews]{Brescia2021,Newman2022}. To systematically evaluate them, the LSST Dark Energy Science Collaboration (DESC) benchmarked twelve \pz codes under controlled conditions in Data Challenge 1 \citep[DC1;][]{Schmidt2020}.

Recently, tabular foundation models (TFMs) pretrained on synthetic datasets have emerged as powerful tools for regression and classification, excelling even in small-sample regimes \citep{Hollmann2023, Qu2025}.
These transformers \citep{Vaswani2017} are pretrained on billions of synthetic tables drawn from random structural causal models \citep[SCMs;][]{Pearl2009} and then perform in-context learning, wherein labeled training examples and unlabeled queries are supplied as input tokens, and targets are predicted in a single forward pass with no weight updates. Such a network approximates the Bayesian posterior predictive distribution under its pretraining prior \citep{Muller2022}. Trained in context, these models can then match task-specific estimators on linear regression, decision trees, and small neural networks \citep{Garg2022}. A TFM thus replaces the fixed inductive bias of a single bespoke estimator with a prior learned over a broad family of data-generating processes.

Recent studies indicate that TFMs can outperform traditional regression methods on selected \pz datasets \citep{Izbicki2026, Valenca2026}. However, they do not fully explore sample efficiency, performance when the training context is not representative of the test set, or the in-context use of a small calibration sample, the three properties an effective learned prior should provide. In this work, we therefore evaluate TFMs, with and without realistic spectroscopic selection bias, on the LSST DESC DC1 \citep{Schmidt2020} benchmark, which simulates Rubin Observatory photometry with a matched spectroscopic training catalog.
We analyze performance scaling with context size, demonstrate that TFMs systematically and substantially outperform bespoke estimators with a fraction of the training data, and show that a small representative calibration sample supplied in context almost entirely mitigates selection bias. We also release our prediction infrastructure as a Python package alongside code to reproduce our results\footnote{\url{https://github.com/biprateep/lazy-tfm}}.

\section{Methods}
\label{methods}

\paragraph{In-context Learning for Conditional Density Estimation.}
To infer \pzs, we estimate the conditional probability density function (PDF), $p(z \mid \mathbf{x})$, where $\mathbf{x} \in \mathbb{R}^D$ denotes observed multi-band photometry and $z \in [0, z_{\rm max}]$ is the true redshift.
Standard supervised estimators are fit to a training catalog $\mathcal{D}_{\rm train} = \{(\mathbf{x}_i, z_i)\}_{i=1}^{N_{\rm train}}$ and must be retrained and re-tuned for each survey.
In contrast, TFMs frame conditional density estimation as in-context learning. Conditioned on a prompt of labeled training context $\mathcal{D}_{\rm train}$ and an unlabeled query galaxy $\mathbf{x}_*$, the predictive PDF $p(z \mid \mathbf{x}_*, \mathcal{D}_{\rm train})$ is obtained in a single forward pass with frozen weights and no gradient updates, fine-tuning, or hyperparameter searches.

\paragraph{Tabular Foundation Models.}
We evaluate three leading\footnote{As of 1st October 2026} open-weight TFMs from the TabArena benchmark \citep{Erickson2025}: Kumo Tabular~\citep{qu2026kumotabular}, LimiX-2~\citep{zhang2026limix2contextualmechanismnetwork, wang2026limix2m} and TabPFN-3.5~\citep{jaeger2026tabpfn35technicalreport}. TabPFN-3.5 and LimiX-2 treat regression as classification over $5{,}000$ buckets of the standardized target, so their native output is the probability in each bucket, while Kumo predicts $999$ quantiles. Taking the density as uniform within each bucket and between consecutive quantiles, both forms give a piecewise-linear cumulative distribution, which we evaluate at the edges of a common grid of $4{,}000$ bins ($\Delta z = 0.0005$) to obtain the probability in each bin without further interpolation. To reduce prediction variance, we ensemble $8$ predictions with different feature permutations and preprocessing. As LimiX-2 was optimized for contexts of up to about $20{,}000$ rows, we use bootstrap aggregating (also called \textit{bagging};~\citealt{Breiman1996}) to use the full dataset. Each ensemble member sees a bootstrap-resampled training context of $15{,}000$ objects, and the predicted densities are averaged.

\paragraph{Baseline Estimators.}
We benchmark against the three top-performing algorithms from the LSST DESC DC1 study \citep{Schmidt2020}: FlexZBoost \citep{Izbicki2017,Dalmasso2020}, Color-Matched Nearest Neighbors \citep[CMNN;][]{Graham2018}, and GPz \citep{Almosallam2016}. FlexZBoost estimates conditional densities by regressing orthonormal cosine basis coefficients using gradient-boosted trees \citep[XGBoost;][]{Chen2016}. CMNN kernel-smooths spectroscopic redshifts of photometric neighbors in color-magnitude space, and GPz models heteroscedastic uncertainties using non-stationary sparse Gaussian processes. All baselines are implemented and evaluated within the Redshift Assessment Infrastructure Layers \citep[RAIL;][]{RailTeam2026} package; newer neural estimators in RAIL are left to future work. We trained the three baseline methods and performed an extensive hyperparameter search with Optuna~\citep{Akiba2019} to determine the optimal settings on the benchmark data (see Appendix~\label{sec:hyperparameters}). Across all metrics, our implementations outperform the results reported in \citet{Schmidt2020}. For a direct comparison with the TFMs, we evaluate all the predicted distributions on the common grid.

\paragraph{Benchmark Datasets and Selection Bias.}
We evaluate all estimators on the LSST DESC DC1 benchmark simulating Rubin Observatory photometry. DC1 is a catalog-level simulation without blending or image-level systematics, so absolute performance is optimistic \citep{Schmidt2020}; as such, we focus on the relative ranking of estimators. Input features comprise the $i$-band magnitude alongside five adjacent colors ($u-g, g-r, r-i, i-z, z-y$) and their associated measurement uncertainties. In the representative regime, where train and test sets are exchangeable, we follow the DC1 configuration with $43{,}486$ training galaxies and $390{,}990$ held-out test galaxies, scored on a $4{,}000$-bin redshift grid across $z \in [0, 2]$. We also evaluate random subsets to quantify scaling with training data size.
To simulate realistic selection bias, we apply the Subaru Hyper Suprime-Cam (HSC) Public Data Release-2 (PDR2) spectroscopic selection function via RAIL \citep[][]{RailTeam2026}, yielding a biased training set of $35{,}011$ galaxies evaluated on a representative held-out test set (see Appendix~\ref{sec:selection_function}).\footnote{See also \citealt{Crafford2026} for a systematic study of training-set degradation with these estimators.}

\paragraph{Evaluation Metrics.}
Following \citet{Schmidt2020}, we extract the point prediction as the distribution mode $z_{\rm peak}$ and evaluate the scaled residual $e_z = (z_{\rm peak} - z_{\rm true}) / (1 + z_{\rm true})$, median residual (bias), normalized interquartile range $\sigma_{\rm IQR} = (Q_{75} - Q_{25}) / 1.349$, and catastrophic outlier fraction $f_{\rm out}$ with threshold $\max(3\sigma_{\rm IQR}, 0.06)$. To evaluate full PDFs, we compute the Conditional Density Estimation (CDE) loss \citep{Izbicki2017}:
\begin{equation}
  \mathcal{L}_{\rm CDE} = \frac{1}{N_{\rm test}} \sum_{i=1}^{N_{\rm test}} \left[ \int p_i^2(z)\,dz - 2\,p_i(z_{{\rm true}, i}) \right],
  \label{eq:cde_loss}
\end{equation}
where $p_i(z)$ is the predicted conditional density for galaxy $i$. The CDE loss is a proper scoring rule that estimates the mean integrated squared error between the estimated and true distributions up to an additive constant, where lower (more negative) values indicate sharper, more accurate distributions. Furthermore, it provides a more reliable measure of calibration than commonly used diagnostics such as the probability integral transform (PIT) and cannot be fooled by pathological examples~\citep{Schmidt2020,Zhao21PIT,Dey2025Ladar}.

\paragraph{Mitigating Selection Bias.}
We follow recent observational strategies that advocate collecting small, representative deep spectroscopic samples \citep[e.g.,][]{Newman2022,Dey2026}, and test concatenating a modest representative calibration sample directly into the in-context prompt. While we tested importance reweighting of the biased training set \citep{Lima2008,Izbicki2017b} and stratified learning \citep[StratLearn;][]{Autenrieth2024} as possible alternatives to direct in-context concatenation, both yielded negligible gains ($<10\%$ relative CDE loss change; cf.\ shift-robust pretraining priors, \citealt{Helli2024}).

\begin{table}[t]
  \centering
  \setlength{\tabcolsep}{4pt}
  \caption{%
    \small Photo-$z$ estimation at full training context size under representative ($N_{\rm train} = 43{,}486$) and selection-biased ($N_{\rm train} = 35{,}011$) regimes, scored on representative held-out test data. Columns report CDE loss ($\mathcal{L}_{\rm CDE}$), point scatter $\sigma_{\rm IQR}$, bias (median residual), in units of $10^{-4}$, and catastrophic outlier fraction $f_{\rm out}$, as a percentage. Values are means over five seeds (at full size the seeds vary only each model's random state); their standard deviations across seeds are at most $0.18$ in $\mathcal{L}_{\rm CDE}$, $0.0002$ in $\sigma_{\rm IQR}$, $2.1\times10^{-4}$ in bias, and $0.3$ percentage points in $f_{\rm out}$. Lower is better for $\mathcal{L}_{\rm CDE}$, $\sigma_{\rm IQR}$, and $f_{\rm out}$; bias is best at zero. Best values in \textbf{bold}. The two sections use different test sets, so CDE loss is only comparable within columns.
   TFMs achieve the best CDE loss and scatter in both regimes and the lowest outlier fraction (tied with CMNN under selection bias), with TabPFN-3.5 best in CDE loss on representative data and LimiX-2 best in CDE loss and scatter under selection bias; the prediction bias is negligibly small ($<2.2\times10^{-3}$) for all methods.
  }
  \label{tab:summary}
  \begin{tabular*}{\textwidth}{@{\extracolsep{\fill}} l @{\hspace{1.2em}\extracolsep{\fill}} rrrr @{\hspace{1.2em}\extracolsep{\fill}} rrrr @{}}
    \toprule
    & \multicolumn{4}{c}{Representative} & \multicolumn{4}{c}{Biased} \\
    \cmidrule(lr){2-5} \cmidrule(lr){6-9}
    Method & $\mathcal{L}_{\rm CDE}$ & $\sigma_{\rm IQR}$ & bias {\scriptsize$[10^{-4}]$} & $f_{\rm out}$ {\scriptsize$[\%]$} & $\mathcal{L}_{\rm CDE}$ & $\sigma_{\rm IQR}$ & bias {\scriptsize$[10^{-4}]$} & $f_{\rm out}$ {\scriptsize$[\%]$} \\
    \midrule
    FlexZBoost & $-10.87$ & $0.0167$ & $-4.5$ & $7.7$ & $-6.66$ & $0.0182$ & $-21.5$ & $11.1$ \\
    CMNN & $-11.23$ & $0.0172$ & $-2.6$ & $6.9$ & $-12.07$ & $0.0172$ & $-16.2$ & $\mathbf{7.0}$ \\
    GPz & $-11.21$ & $0.0182$ & $\mathbf{0.2}$ & $8.9$ & $-10.62$ & $0.0202$ & $-2.2$ & $12.0$ \\
    TabPFN-3.5 & $\mathbf{-15.30}$ & $0.0138$ & $-1.1$ & $6.1$ & $-15.16$ & $0.0149$ & $-6.8$ & $\mathbf{7.0}$ \\
    LimiX-2 & $-15.15$ & $\mathbf{0.0135}$ & $\mathbf{0.2}$ & $6.1$ & $\mathbf{-15.26}$ & $\mathbf{0.0143}$ & $-4.1$ & $7.2$ \\
    Kumo & $-14.95$ & $0.0140$ & $0.8$ & $\mathbf{6.0}$ & $-14.97$ & $0.0148$ & $\mathbf{1.1}$ & $7.1$ \\
    \bottomrule
  \end{tabular*}
\end{table}
\section{Results and Discussions}
\label{results}

\begin{figure}[t]
  \centering
 
  \includegraphics[width=\textwidth]{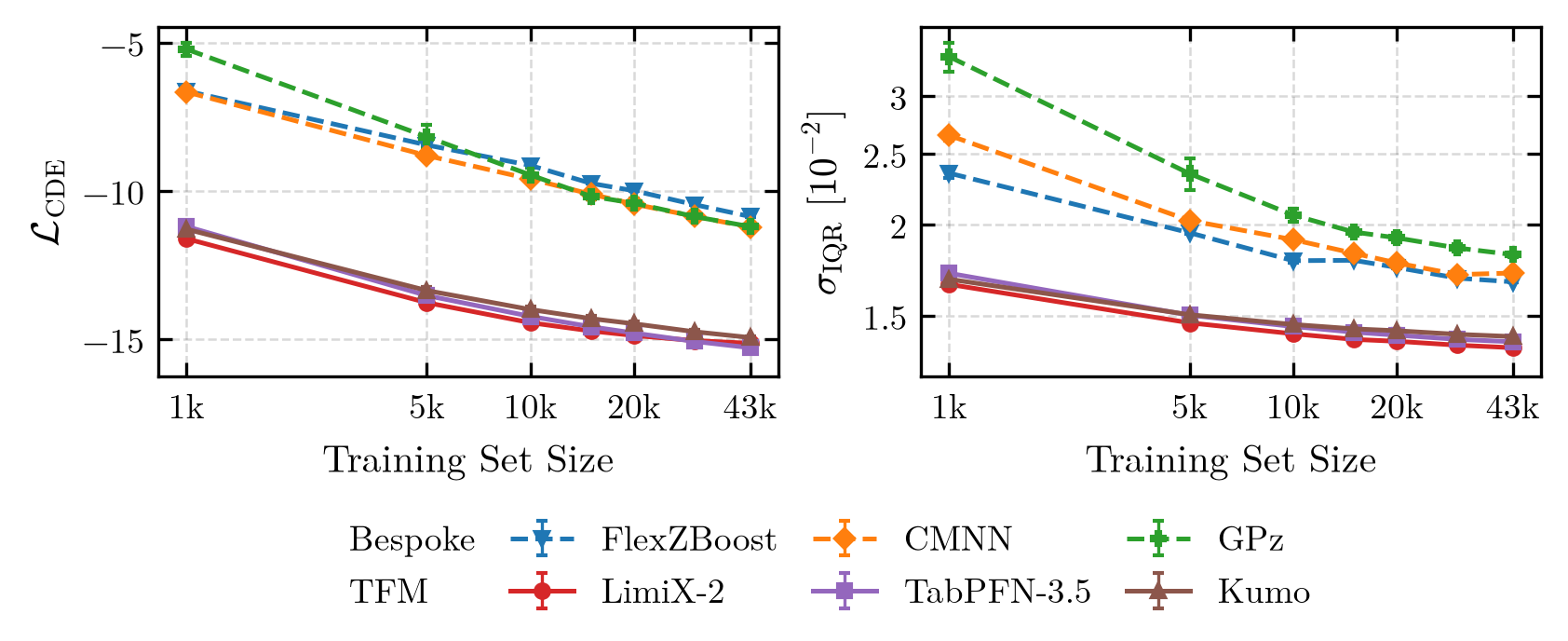}
  \caption{%
    \small Performance scaling of TFMs and bespoke \pz estimators on the representative DC1 benchmark across training context sizes.
    \textbf{Left}: CDE loss ($\mathcal{L}_{\rm CDE}$, lower is better).
    \textbf{Right}: Point dispersion $\sigma_{\rm IQR}$ at distribution mode $z_{\rm peak}$ (log scale).
   TFMs continue to improve out to the full catalog. With just $1{,}000$ context galaxies, all three TFMs match or outperform full-catalog bespoke baselines, and by $5{,}000$ they substantially outperform all of them.
  }
  \label{fig:unbiased_scaling}
\end{figure}

Table~\ref{tab:summary} compares all six estimators at full context size on representative ($N_{\rm train} = 43{,}486$) and biased ($N_{\rm train} = 35{,}011$) DC1 data. On representative data, all three TFMs outperform bespoke estimators in CDE loss, scatter, and outlier fraction. TabPFN-3.5 attains the best CDE loss ($\mathcal{L}_{\rm CDE} = -15.30$), followed by LimiX-2 ($-15.15$) and Kumo ($-14.95$), while LimiX-2 has the lowest scatter ($\sigma_{\rm IQR} = 0.0135$). Compared to the strongest bespoke baselines (CMNN at $\mathcal{L}_{\rm CDE} = -11.23$ on par with GPz at $-11.21$), TabPFN-3.5 improves CDE loss by $4.07$ units, reduces scatter by $\sim 20\%$, and lowers the outlier fraction by $0.8$ percentage points. Figure~\ref{fig:unbiased_scaling} shows the performance on the representative training set as a function of training context size, and we see that all three TFMs with just $1{,}000$ context galaxies ($\mathcal{L}_{\rm CDE} = -11.62$ to $-11.21$) match or outperform all three bespoke estimators trained on the full $43{,}486$ catalog ($\mathcal{L}_{\rm CDE} = -11.23$ to $-10.87$).

Under spectroscopic selection bias (Table~\ref{tab:summary}, right), all three TFMs reach a lower CDE loss than every bespoke estimator trained on the same biased catalog, with LimiX-2 at $\mathcal{L}_{\rm CDE} = -15.26$ and $\sigma_{\rm IQR} = 0.0143$ vs.\ $-12.07$ and $0.0172$ for CMNN, the strongest bespoke method in this regime. The TFMs fall within $0.29$ units of CDE loss of one another ($\mathcal{L}_{\rm CDE} = -15.26$ to $-14.97$), whereas the bespoke estimators spread over $5.41$ units ($\mathcal{L}_{\rm CDE} = -12.07$ to $-6.66$). All three TFMs also keep a lower scatter than every bespoke estimator ($\sigma_{\rm IQR} = 0.0143$ to $0.0149$ vs.\ $0.0172$ to $0.0202$). This is consistent with the TFMs not extrapolating catastrophically into regions of feature space left uncovered by the biased training set. Scaling trends are similar to the representative case (see Appendix Figure~\ref{fig:biased_scaling}), except that with $1{,}000$ biased galaxies the TFMs ($\mathcal{L}_{\rm CDE} = -11.74$ to $-11.45$) outperform GPz and FlexZBoost trained on the full biased catalog but not CMNN ($-12.07$), which all three overtake by $5{,}000$ galaxies ($-13.93$ to $-13.68$).
\begin{figure}[t]
  \centering
  \includegraphics[width=\textwidth]{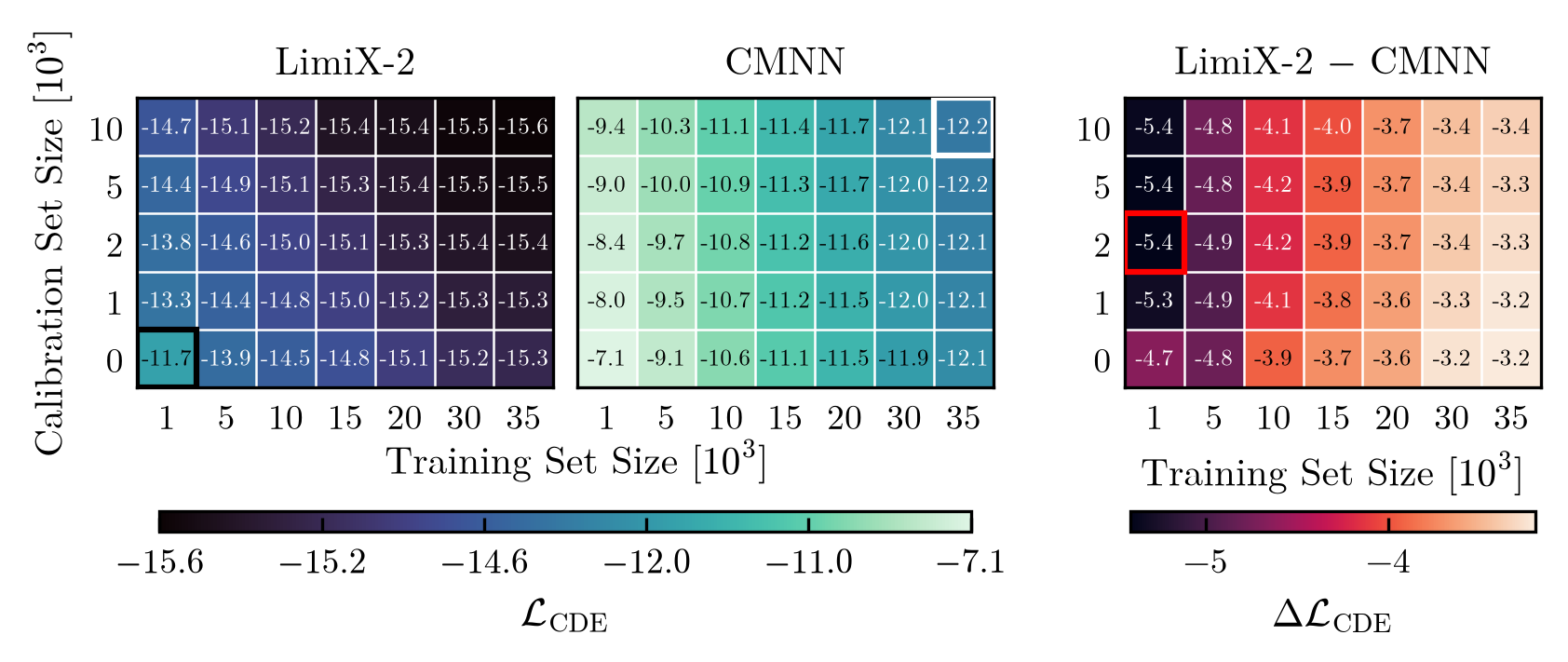}
  \caption{%
    \small CDE loss of LimiX-2 (\textbf{left}) and of CMNN (\textbf{middle}), the strongest bespoke estimator at full biased context, across biased training ($N_{\rm train} = 1{,}000$ to $35{,}011$) and representative calibration sample sizes ($N_{\rm cal} = 0$ to $10{,}000$), on a shared color scale, along with their difference (\textbf{right}) $\Delta\mathcal{L}_{\rm CDE} = \mathcal{L}_{\rm LimiX} - \mathcal{L}_{\rm CMNN}$. Lower is better throughout. The outlined cells in the left and middle panels mark the least data given to LimiX-2 ($1{,}000$ biased galaxies and no calibration sample) and the most given to CMNN ($35{,}011$ biased galaxies and $10{,}000$ calibration galaxies), and the cell with the red border in the right panel marks where the advantage of LimiX-2 is largest ($\Delta\mathcal{L}_{\rm CDE} = -5.4$, with $1{,}000$ biased and $2{,}000$ calibration galaxies). LimiX-2 outperforms CMNN in all $35$ cells, by $3.2$ to $5.4$ units. With the least data, LimiX-2 trails CMNN with the most ($-11.74$ vs.\ $-12.23$), but with $1{,}000$ biased galaxies and $1{,}000$ calibration galaxies, LimiX-2 ($-13.35$) already outperforms CMNN with $35{,}011$ biased galaxies and $10{,}000$ calibration galaxies ($-12.23$).
    }
  \label{fig:matrix_tabfm}
\end{figure}

Figure~\ref{fig:matrix_tabfm} shows the CDE loss of LimiX-2 and of CMNN, the strongest bespoke estimator under a non-representative training set, across context ($N_{\rm train} = 1{,}000$ to $35{,}011$) and calibration data sizes ($N_{\rm cal} = 0$ to $10{,}000$). While representative calibration galaxies improve both, the size of the gain depends on how much non-representative context the model already has. With $1{,}000$ training objects, $1{,}000$ calibration galaxies lower LimiX-2's CDE loss by $1.61$ units ($\mathcal{L}_{\rm CDE} = -11.74$ to $-13.35$), whereas at the full $35{,}011$ training context, the same addition gains only $0.08$ units ($-15.26$ to $-15.34$). Unlike the representative case, LimiX-2 with $1{,}000$ biased galaxies and no calibration sample trails CMNN with the full non-representative training set and $10{,}000$ additional calibration galaxies ($\mathcal{L}_{\rm CDE} = -11.74$ vs.\ $-12.23$). In that configuration it still matches or slightly outperforms GPz ($-11.42$) and clearly outperforms FlexZBoost ($-8.29$) when these are given the same full biased training set and $10{,}000$ calibration galaxies. However, adding $1{,}000$ calibration galaxies ($-13.35$) or using $5{,}000$ training galaxies alone ($-13.93$) is enough for LimiX-2 to overtake CMNN's best configuration, and the same holds for TabPFN-3.5 ($-13.01$ and $-13.69$) and Kumo ($-13.09$ and $-13.68$; see Appendix~\ref{sec:biased_performance}). These results suggest that directly appending representative training data to the prompt is an effective and sample-efficient strategy. The TFMs do not gain the most from it (with $1{,}000$ biased galaxies, $1{,}000$ calibration galaxies lower the loss by $1.61$ units for LimiX-2 vs.\ $0.95$ for CMNN, $1.82$ for FlexZBoost, and $1.90$ for GPz), but they keep their lead at every combination of biased and calibration sample sizes. Because faint spectroscopy is expensive, modest deep-field samples ($N_{\rm cal} \sim 1{,}000$) should be a cost-effective way to restore parameter-space coverage, with the largest gains where the biased context is smallest.

In summary, TFMs deliver exceptionally accurate in-context \pzs without survey-specific retraining. With $1{,}000$ context examples they match bespoke estimators trained on $43{,}486$ representative galaxies, and under spectroscopic selection bias they lead every bespoke estimator trained on the same biased and calibration samples by at least $2.90$ units in CDE loss. With only $1{,}000$ biased galaxies, however, they need a small representative calibration sample to overtake CMNN trained on the full biased catalog. On DC1, the TFM point estimates at full context size meet the per-galaxy LSST targets quoted by \citet{Schmidt2020}. While meeting the DESC requirements on tomographic $n(z)$ \citep{TheLSSTDarkEnergyScienceCollaboration2018} requires a dedicated calibration analysis on realistic data, their strong performance under realistic conditions leads us to advocate for TFMs as drop-in \pz estimators in future studies, with fine-tuning and calibration as future refinements rather than prerequisites.

\clearpage
\begin{ack}
The authors thank Tristan Fattedad for insightful discussions and for conducting initial prototyping experiments on a related problem that inspired this work.

B. Dey is a postdoctoral fellow at the University of Toronto in the Eric and Wendy Schmidt AI in Science Postdoctoral Fellowship Program, a program of Schmidt Sciences.

This research used resources of the National Energy Research Scientific Computing Center (NERSC), a Department of Energy User Facility using NERSC award HEP-ERCAP0036802.

The authors used Anthropic's Claude Opus 5.5 via the Claude Code harness to assist in developing the Python code for the experiments, and Google's Gemini 3.8 Flash to edit an initial human-written draft for spelling, grammar, and conciseness to meet the workshop page limits. The authors are solely responsible for the ideation, experimental design, and interpretation of the results, and retain full responsibility for the accuracy, integrity, and content of this work.
\end{ack}

\bibliographystyle{plainnat}
\bibliography{references}

\clearpage
\appendix

\section{Technical appendices and supplementary material}

\subsection{Spectroscopic Selection Function and Sample Properties}
\label{sec:selection_function}

\begin{figure}[!ht]
  \centering
  \includegraphics[width=\textwidth]{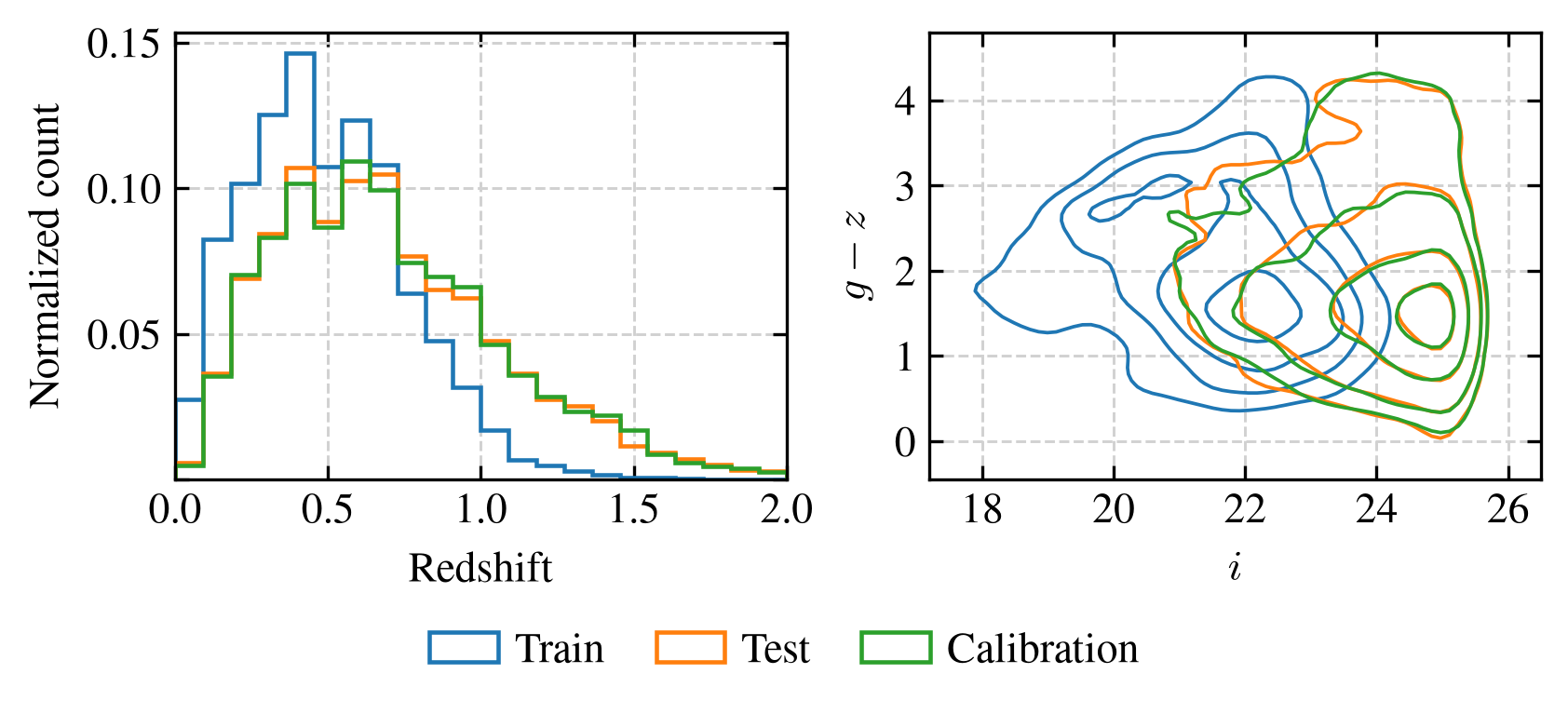}
  \caption{%
    \small Impact of the spectroscopic selection function on the training sample distributions compared to the representative test and calibration samples.
    \textbf{Left}: Redshift distribution showing the biased spectroscopic training sample (blue), the representative test sample (orange), and the representative calibration sample (green).
    \textbf{Right}: Distribution in the selection plane of $i$-band magnitude vs.\ $(g - z)$ color, where solid contours enclose 11.8\%, 39.3\%, 67.5\%, and 86.4\% of each sample's galaxies (the two-dimensional Gaussian equivalents of $0.5\sigma$, $1\sigma$, $1.5\sigma$, and $2\sigma$).
    We observe that the spectroscopic selection preferentially removes faint ($i > 24$) and high-redshift ($z > 0.8$) galaxies, producing an unrepresentative training distribution, whereas the calibration sample closely matches the representative test population across all magnitudes and colors.
  }
  \label{fig:selection_samples}
\end{figure}

In real spectroscopic surveys, finite exposure times and target prioritization preferentially yield redshifts for bright galaxies with prominent emission or absorption features. This induces both covariate shift in optical photometry and label shift in redshift. To model this observational reality within the LSST DESC DC1 benchmark, we implement a spectroscopic selection algorithm based on the Subaru Hyper Suprime-Cam (HSC) PDR2 spectroscopic sample, adapted from the Redshift Assessment Infrastructure Layers \citep[RAIL;][]{RailTeam2026}.

To construct the biased training sample, we apply a two-dimensional selection over $i$-band magnitude and $(g - z)$ color combined with a color-dependent redshift ceiling. We first map each galaxy onto a $100 \times 100$ grid spanning $13 \le i \le 26$ and $-2 \le g - z \le 6$ and assign it the spectroscopic success ratio $r$ (defined as the ratio of spectroscopic to photometric galaxies in HSC PDR2 and available as a downloadable dataset) of its corresponding cell, setting $r = 0$ outside the grid boundaries. We then compute the 99th percentile of HSC spectroscopic redshifts observed in each cell and discard any galaxy whose redshift exceeds this ceiling, removing all galaxies from cells lacking spectroscopic coverage. This step is critical because it makes the selection explicitly dependent on redshift rather than photometry alone, introducing a truncation that cannot be inverted from photometric features. Finally, within each distinct ratio cell, galaxies are retained at random with probability proportional to $r$.

Because this selection retains only $8.66\%$ of representative galaxies, the full $434{,}476$-galaxy DC1 parent catalog yields at most $37{,}626$ biased candidates. To ensure sufficient holdout data for evaluation, we carve out a biased training pool of $35{,}011$ galaxies. From the remaining holdout, we construct an independent representative test set of $19{,}383$ galaxies and a representative calibration sample of $10{,}000$ galaxies.

Figure~\ref{fig:selection_samples} shows the resulting distributions of the biased training, representative test, and calibration samples in redshift and color-magnitude space. In the left panel, we see that the spectroscopic selection severely alters the redshift distribution: the biased training sample peaks at $z \sim 0.4$ and cuts off sharply above $z \sim 0.8$, with fewer than $1\%$ of galaxies retained at $z > 1.2$. In contrast, the representative test and calibration samples extend continuously out to $z = 2.0$. The right panel shows the sample distributions in the $(i, g - z)$ selection plane, where solid contours enclose 11.8\%, 39.3\%, 67.5\%, and 86.4\% of each sample. We observe that the biased training sample is heavily concentrated at bright magnitudes ($i \lesssim 22.5$), leaving the faint regime ($i \gtrsim 24$) almost entirely unpopulated. Meanwhile, the calibration sample closely tracks the representative test distribution across both bright and faint magnitudes, illustrating how including even a modest representative calibration sample supplies the missing support across color-magnitude space.

\subsection{Performance under Selection Bias}
\label{sec:biased_performance}

\begin{figure}[t]
  \centering
  \includegraphics[width=\textwidth]{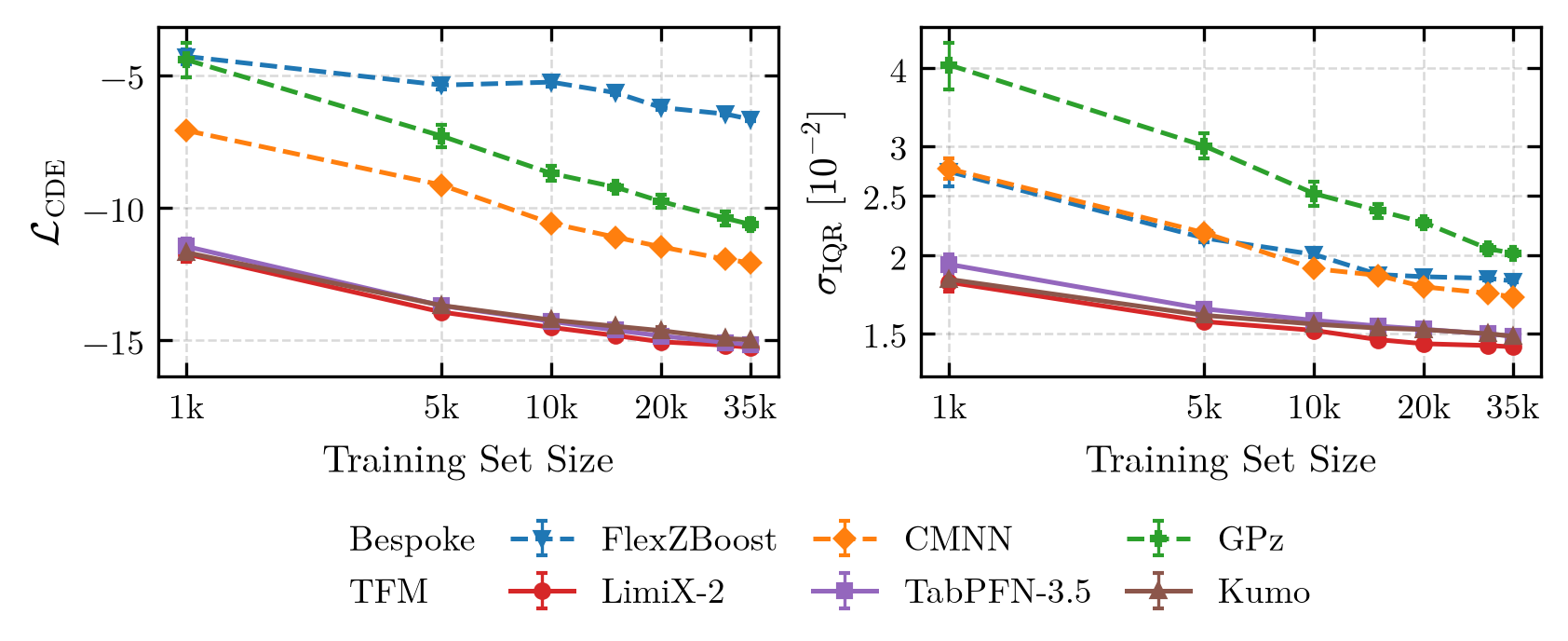}
  \caption{%
    \small As Figure~\ref{fig:unbiased_scaling}, but showing scaling of TFMs and bespoke \pz estimators under realistic spectroscopic selection bias as a function of training data size, evaluated on the same test set.
    We observe that TFMs scale steadily with training context size and outperform the bespoke estimators in both metrics at every training size, with LimiX-2 achieving the lowest CDE loss and scatter across all training sizes. In contrast, FlexZBoost, which performed well on the representative training data, does not scale well on the biased sample. All TFMs achieve better CDE loss with just $1{,}000$ training examples compared to GPz and FlexZBoost with the full data, and with $5{,}000$ training examples compared to CMNN with the full data.
  }
  \label{fig:biased_scaling}
\end{figure}

Figure~\ref{fig:biased_scaling} shows the performance scaling of all six estimators when trained on the spectroscopically biased dataset ($N_{\rm train} = 1{,}000$ to $35{,}011$) and evaluated on the representative test catalog. As shown in the left panel, the bespoke baselines spread over a wide range of CDE loss under spectroscopic selection, whereas the TFMs remain close together. In particular, FlexZBoost exhibits the flattest scaling curve with $\mathcal{L}_{\rm CDE}$ between  $-4.3$ and $-6.7$, suggesting that gradient-boosted trees struggle to extrapolate into truncated regions of feature space. CMNN and GPz scale more steadily, reaching $\mathcal{L}_{\rm CDE} = -12.07$ and $-10.62$ at full context, respectively. In contrast, all three TFMs scale smoothly toward CDE losses below $-14.9$, with LimiX-2 achieving $-15.26$ at $35{,}011$ galaxies. Remarkably, TFMs trained on just $1{,}000$ biased context galaxies attain CDE losses between $-11.5$ and $-11.7$, outperforming GPz and FlexZBoost trained on the full $35{,}011$-galaxy catalog; only CMNN ($-12.07$) stays ahead, and all three TFMs overtake it by $5{,}000$ galaxies ($-13.7$ to $-13.9$).

The right panel shows the corresponding scaling of point estimate dispersion ($\sigma_{\rm IQR}$). We see that TFMs consistently produce lower scatter across the entire context range. At small training sizes ($N_{\rm train} = 1{,}000$), bespoke estimators suffer substantial dispersion ($\sigma_{\rm IQR} \approx 0.027$ for FlexZBoost and CMNN), whereas the TFMs maintain $\sigma_{\rm IQR} \approx 0.018$ to $0.019$. At full context, LimiX-2 achieves $\sigma_{\rm IQR} = 0.0143$, and Kumo and TabPFN-3.5 $0.0148$ and $0.0149$, outperforming CMNN ($0.0172$), FlexZBoost ($0.0182$), and GPz ($0.0202$). This suggests that the pretraining phase equips TFMs with inductive priors that prevent catastrophic degradation when spectroscopic training samples are truncated.

\begin{figure}[!ht]
  \centering
  \includegraphics[width=\textwidth]{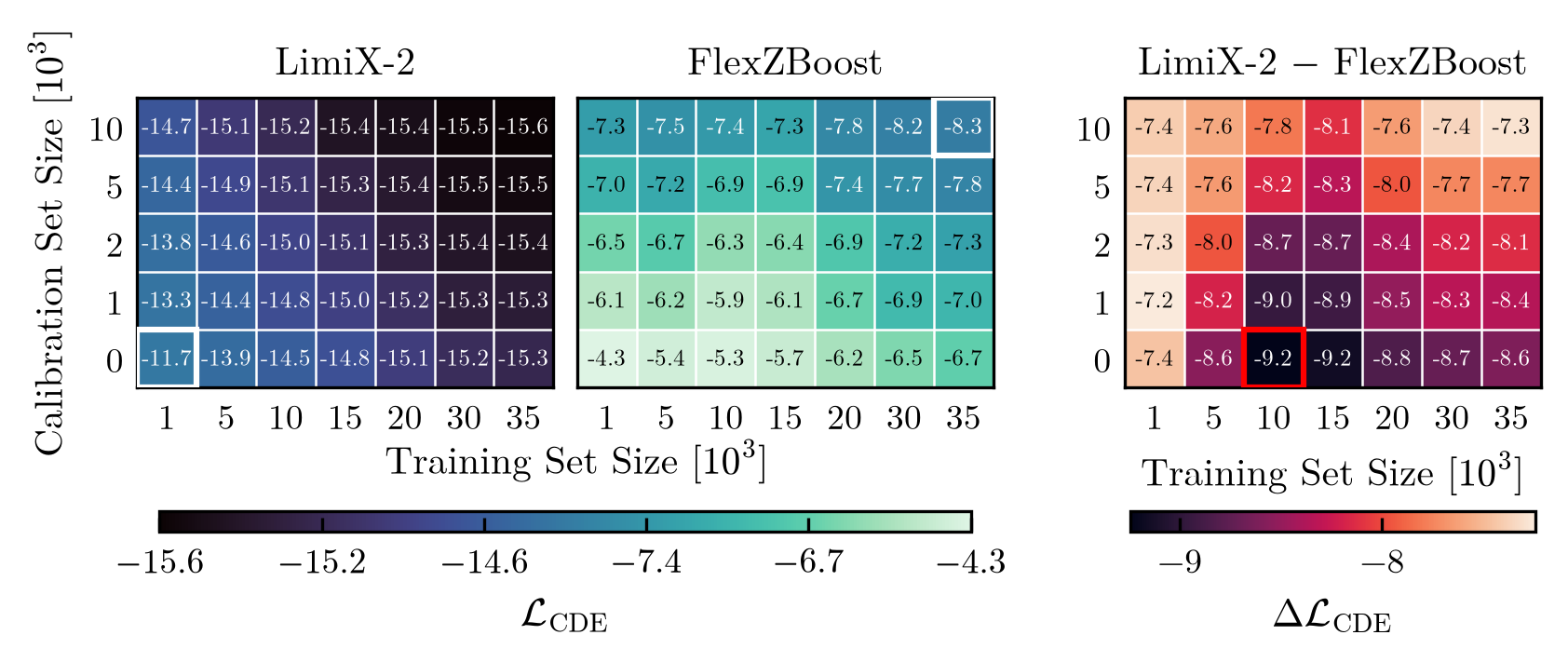}
  \caption{%
    \small As Figure~\ref{fig:matrix_tabfm}, but comparing LimiX-2 to FlexZBoost instead of CMNN. As FlexZBoost has the highest CDE loss of the three bespoke estimators under selection, the margin here is the widest.
  }
  \label{fig:matrix_limix_flexzboost}
\end{figure}

\begin{figure}[!ht]
  \centering
  \includegraphics[width=\textwidth]{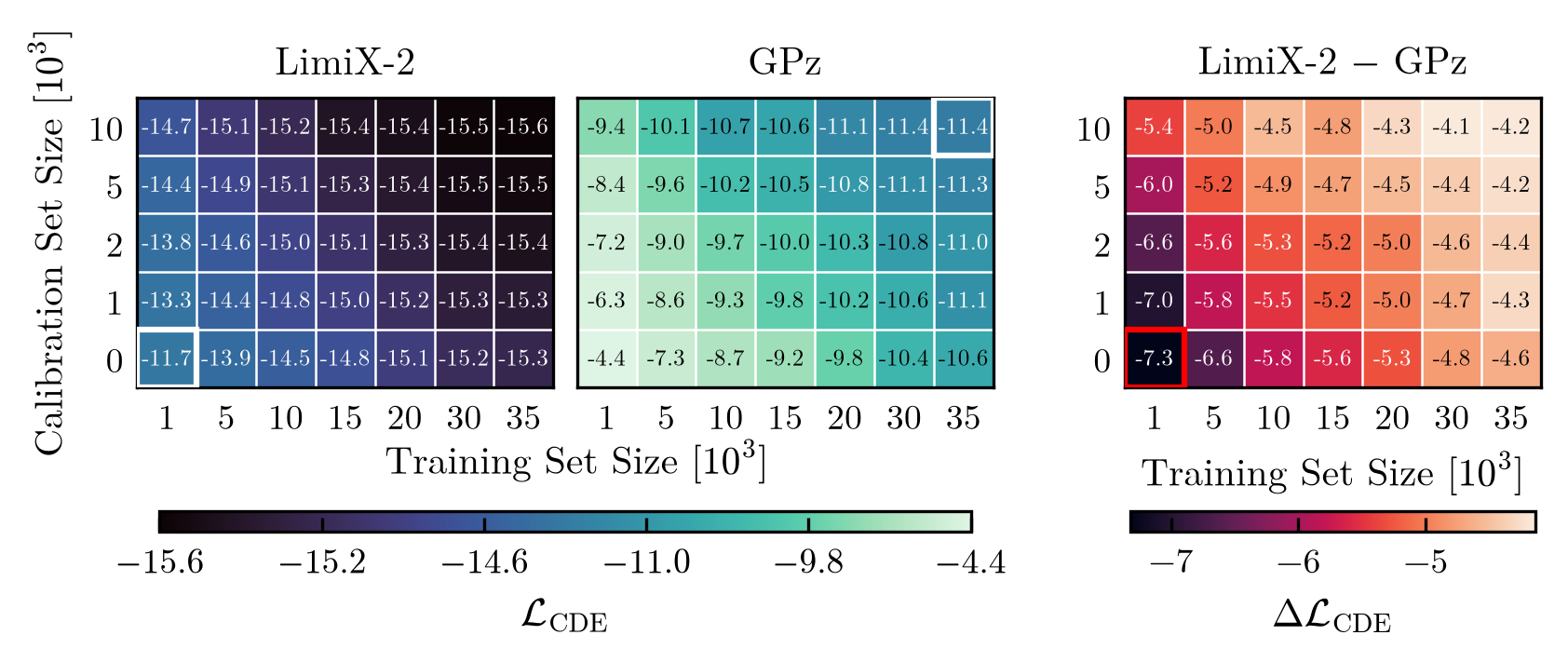}
  \caption{%
    \small As Figure~\ref{fig:matrix_tabfm}, but comparing LimiX-2 to GPz instead of CMNN. At every training set size, the relative advantage of LimiX-2 is smaller with $10{,}000$ calibration galaxies than with none, unlike with CMNN.
  }
  \label{fig:matrix_limix_gpz}
\end{figure}

\begin{figure}[!ht]
  \centering
  \includegraphics[width=\textwidth]{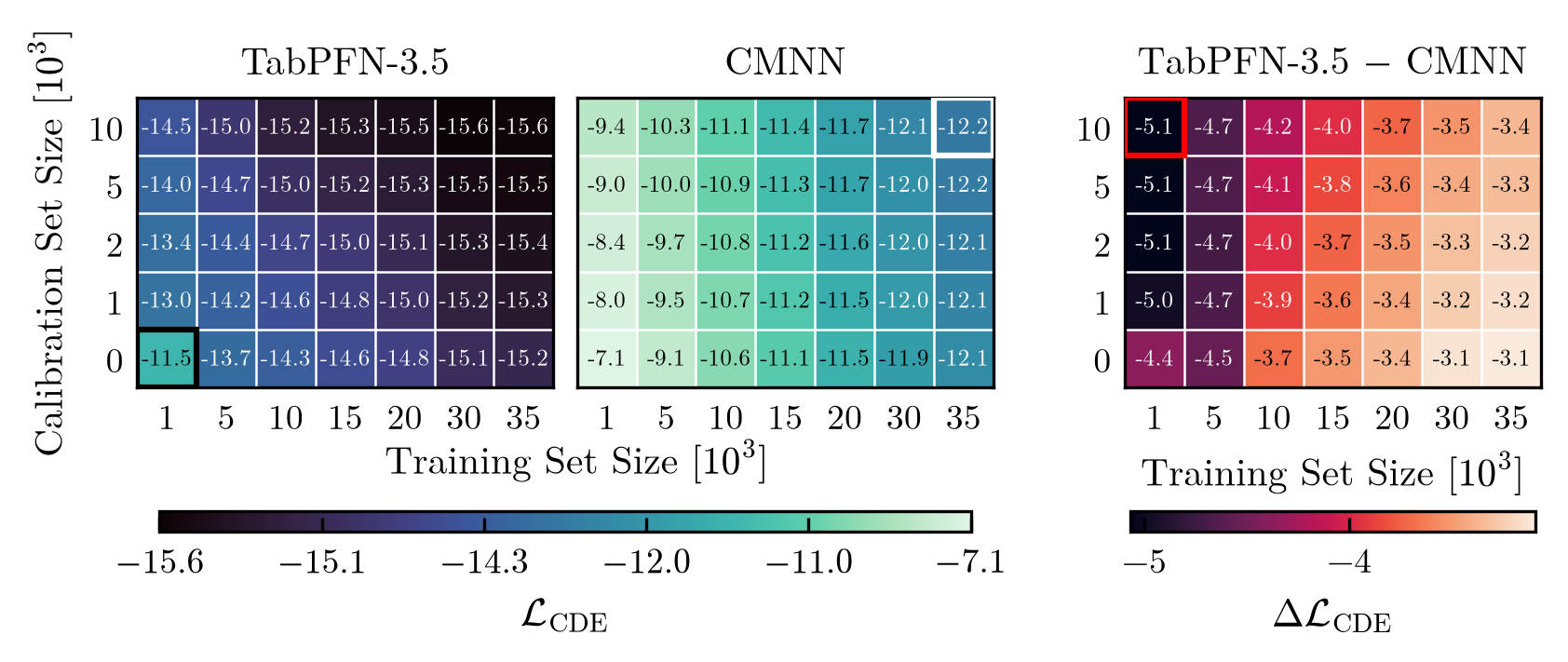}
  \caption{%
    \small As Figure~\ref{fig:matrix_tabfm}, but for TabPFN-3.5 instead of LimiX-2.
  }
  \label{fig:matrix_tabpfn}
\end{figure}

\begin{figure}[!ht]
  \centering
  \includegraphics[width=\textwidth]{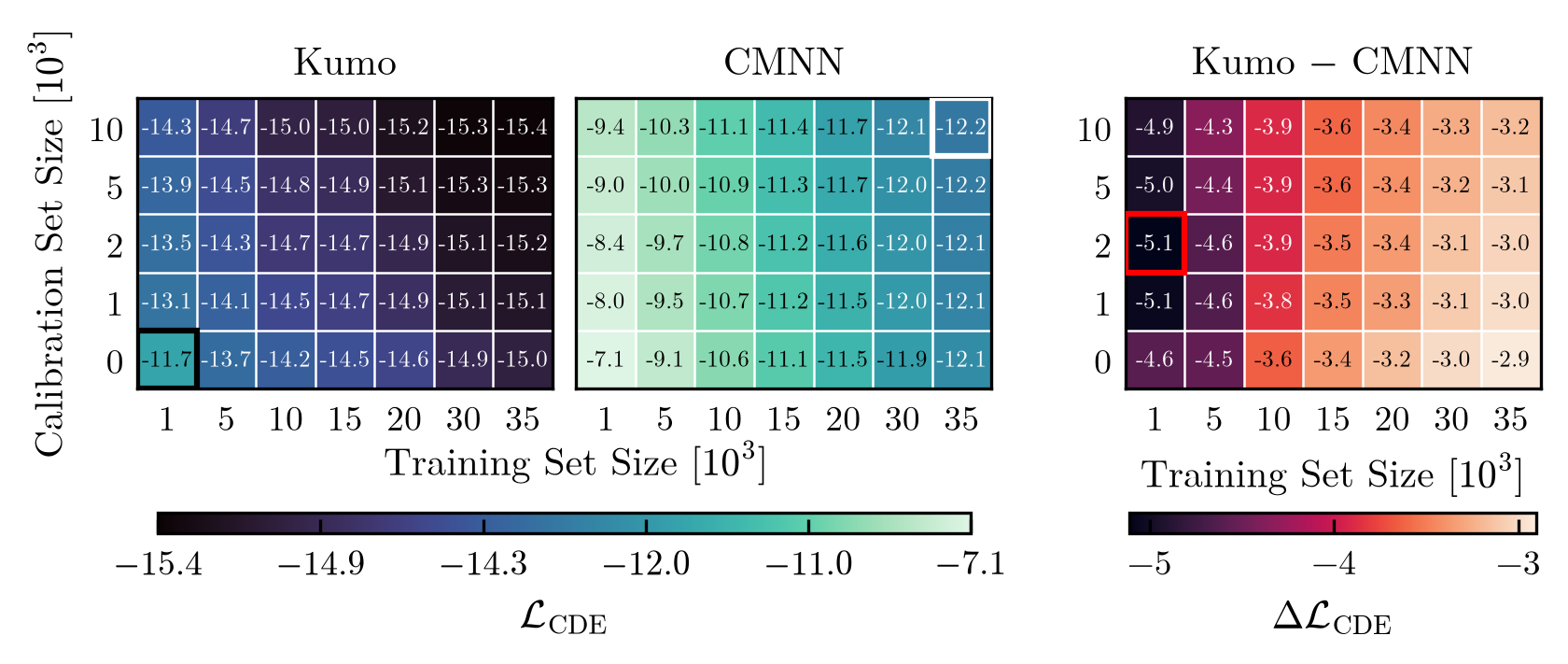}
  \caption{%
    \small As Figure~\ref{fig:matrix_tabfm}, but for Kumo instead of LimiX-2.
  }
  \label{fig:matrix_kumo}
\end{figure}

Figures~\ref{fig:matrix_limix_flexzboost} and \ref{fig:matrix_limix_gpz} show LimiX-2 against the two remaining bespoke baselines, and Figures~\ref{fig:matrix_tabpfn} and \ref{fig:matrix_kumo} show TabPFN-3.5 and Kumo against CMNN, over the same grid of biased context size ($N_{\rm train} = 1{,}000$ to $35{,}011$) and representative calibration sample size ($N_{\rm cal} = 0$ to $10{,}000$). In all cases, each TFM reaches a lower CDE loss than each bespoke estimator under every evaluated combination of biased context and calibration data. The margins are similar across the three TFMs, with LimiX-2 achieving values lower than FlexZBoost by $7.24$ to $9.25$ units and lower than GPz by $4.14$ to $7.32$ units. Similarly, TabPFN-3.5 and Kumo outperform CMNN by $3.09$ to $5.07$ and $2.90$ to $5.11$ units, respectively, closely matching the $3.19$ to $5.41$ range of LimiX-2 (Figure~\ref{fig:matrix_tabfm}). As representative calibration galaxies are added along each column, FlexZBoost and GPz recover faster than LimiX-2 (i.e., the differences between their CDE losses shrink), whereas CMNN recovers more slowly than the TFMs at most training set sizes; however, the TFMs continue to stay ahead in every cell. In-context calibration is therefore a general mechanism across TFMs rather than a property of one backbone.

\subsection{Hyperparameter Search and Final Settings}
\label{sec:hyperparameters}

We tuned every bespoke method separately for each regime and training set size with Optuna~\citep{Akiba2019}, using its multivariate tree-structured Parzen estimator sampler. Each trial fits a method on four folds of the training set and predicts the fifth, repeating this across all five folds, and a trial that completes all five is scored by the CDE loss on the common grid averaged over the held-out folds. A median pruner stops unpromising trials early, and the test set is never used during tuning. Settings that only post-process a fitted model's predictions are evaluated on a grid within every trial, including the number of trees, the number of series terms kept, and the sharpening exponent for FlexZBoost, the kernel width for CMNN, and the scale and floor of the predicted widths for GPz.

On representative data, FlexZBoost uses a Fourier basis of $60$ terms whose coefficients are regressed together by trees with vector leaves ($191$ trees, a learning rate of $0.0068$, and a maximum depth of $13$) and a sharpening exponent of $1.225$. On biased data, it keeps the Fourier basis with $55$ terms and uses $256$ trees, a learning rate of $0.019$, a maximum depth of $16$, and a sharpening exponent of $1.375$. CMNN matches neighbors in the five colors and the $i$-band magnitude, with $0.68$\,mag added in quadrature to the $i$-band error, within the $0.757$ percent point of the $\chi^2$ distribution and with at least $25$ neighbors, weighting them by inverse distance to the power $1.84$ with a distance floor of $4.7$ and smoothing their redshifts with a Gaussian kernel of fixed width $0.0080$. On biased data it adds $0.83$\,mag to the $i$-band error and uses the $0.724$ percent point, at least $15$ neighbors, a power of $1.69$, a floor of $5.1$, a kernel width of $0.8$ times the spread of the matched subset, and the color pre-cut. GPz uses $51$ basis functions, each with its own full covariance, heteroscedastic noise, and the $i$-band magnitude, colors and logarithms of the uncertainties as inputs; it holds out $10\%$ of the training rows to stop the optimization, keeps the best of five restarts, and scales its predicted widths by $0.95$ with a floor of $0.005$ added in quadrature. On biased data it uses $96$ basis functions, decorrelates its inputs with PCA, holds out $5\%$ of the rows, and uses a width scale of $1.0$ with a floor of $0.0025$.

\end{document}